# Engineering Near-Infrared Two-Level Systems in Confined Alkali Vapors

Gilad Orr[1], Golan Ben-Ari[2], Eliran Talker[2]

[1]Department of Physics, Ariel University, Ariel 40700

[2]Department of Electrical Engineering, Ariel University, Ariel 40700, Israel

*Corresponding author: elirant@ariel.ac.il

## Abstract

We combined experimental and theoretical investigations of an effective two-level atomic system operating in the near-infrared telecom wavelength regime, realized using hot rubidium vapor confined within a sub-micron-thick cell. In this strongly confined geometry, atomic coherence is profoundly influenced by wall-induced relaxation arising from frequent atom–surface collisions. By analyzing both absorption and fluorescence spectra, we demonstrate that the optical response is dominated by a closed cycling transition, which effectively isolates the atomic dynamics to a two-level configuration despite the presence of multiple hyperfine states. This confinement-induced selection suppresses optical pumping into uncoupled states and enables robust, controllable light–matter interaction at telecom wavelengths within a miniature atomic platform. Our results establish a practical route to realizing near-infrared atomic two-level systems in compact vapor-cell devices, opening new opportunities for integrated quantum photonic technologies, including on-chip quantum memories, telecom-band frequency references, and scalable quantum information processing.

## Introduction

Spectroscopy between excited atomic states has been widely employed across a broad range of applications, including laser cooling[1,2], high-resolution spectroscopy[3–6], and the development of frequency standards[7]. Among these systems, the $5P_{3/2} \rightarrow 4D_{5/2}$ transition in rubidium has attracted particular attention as a promising atomic reference in the telecom wavelength regime, with potential applications in quantum memories, fiber-based quantum communication[8–13], and quantum-enhanced sensing[14–17].

Accessing excited-state transitions in alkali vapor typically relies on optical–optical double-resonance (OODR) spectroscopy in a ladder-type atomic configuration, which enables Doppler-free measurements of transitions between excited states. Sub-Doppler OODR spectra of the $5P_{3/2} \rightarrow 4D_{5/2}$ transition have been demonstrated in rubidium vapor cells; however, the hyperfine structure of the $4D_{5/2}$ manifold in Rb has remained largely unresolved. This limitation arises from the small hyperfine splittings of the excited state and from the fact that the cycling transition dominates the OODR signal, overwhelming weaker hyperfine components. Moreover, the signal-to-noise ratio (SNR) of excited-state OODR spectroscopy is inherently limited: the short lifetime of the intermediate $5P_{3/2}$ state (~ 26 ns), compared to the longer lifetime of the $4D_{5/2}$ state (~ 84 ns), leads to reduced population transfer and consequently weak excited-state signals, even for the cycling transition.

To overcome these limitations, Moon *et al*[18–20] introduced the double-resonance optical pumping (DROP) technique, which enables high-SNR spectroscopy of excited-state transitions by detecting changes in the ground-state population rather than directly probing the excited-state population. In this approach, atoms are optically pumped through the intermediate and excited states into an uncoupled ground state, producing a strong and spectrally resolved signal. Building on this concept, we recently developed fluorescence double-resonance optical pumping (FDROP)[21], in which the excited-state population dynamics are inferred from fluorescence rather than ground-state transmission. For millimeter-scale vapor cells, FDROP provides an improved SNR compared to conventional DROP spectroscopy. Nevertheless, in both DROP and FDROP, the dominant spectral features arise from transitions that pump atoms into uncoupled ground states. As a result, non-cycling transitions dominate the spectra, which fundamentally limits the realization of an effective two-level system.

Ultrathin alkali-vapor cells provide a powerful route to overcoming this limitation. Sub-Doppler spectroscopy in ultrathin cells originates from the anisotropic atom–light interaction imposed by wall-to-wall trajectories: atoms with slow velocity components normal to the cell windows experience longer interaction times and therefore contribute disproportionately to the detected signal[22–36]. Under normal-incidence illumination, these slow atoms travel nearly parallel to the cell walls, rendering their resonances effectively insensitive to Doppler shifts. This velocity-selective mechanism naturally suppresses Doppler broadening and optical pumping into unwanted states.

In this work, we investigate excited-state spectroscopy in the near-infrared telecom regime using ultrathin rubidium vapor cells. We show that for cell thickness below 5 µm, the dynamics of slow atoms strongly favor the cycling $5P_{3/2} \rightarrow 4D_{5/2}$ transition. By reconstructing the atomic density-matrix populations, we demonstrate that the cycling transition becomes dominant in this confined geometry, enabling the realization of an effective near-infrared two-level system. These results open new pathways toward compact telecom-band quantum memories, atomic-referenced laser frequency stabilization, fiber-based quantum communication, and quantum-enhanced sensing applications such as gyroscopes and magnetometers.

The paper is organized as follows: we first present the theoretical model and numerical simulations, followed by the experimental setup and measurements, and conclude with a discussion of the implications of our results.

## Theory

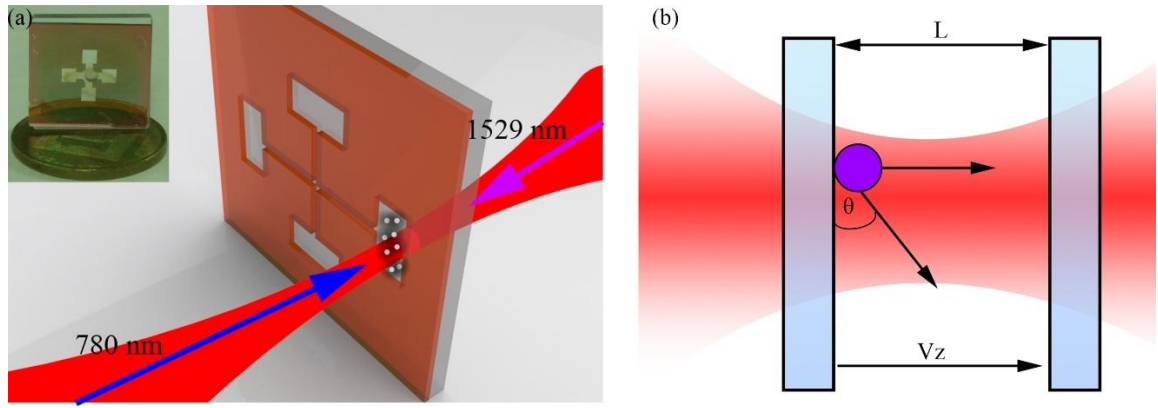

**FIG. 1. Ultrathin vapor cell geometry and velocity-selective atomic trajectories** (a) Schematic of the homemade ultrathin vapor cell. Counter-propagating probe (780 nm) and coupling (1529 nm) beams propagate through cells of different thicknesses (500 nm, 1 µm, 5 µm, and 30 µm), as indicated in the squares. Inset: photograph of the fabricated thin cell. (b) Illustration of atomic trajectories inside the thin cell. Atoms with small longitudinal velocity ($v_z$), traveling nearly perpendicular to the cell windows, experience a significantly longer light–matter interaction time compared to atoms with larger ($v_z$), which traverse the cell more rapidly.

The preparation of a near-infrared two-level system (TLS) using hot atomic vapor confined in a homemade thin cell is illustrated in Fig. 1(a). Two counter-propagating probe and coupling beams intersect within the thin cell. As shown in Fig. 1(b), this geometry gives rise to a velocity-selective process, in which only atoms with small longitudinal velocities contribute significantly to the transmitted and fluorescence signals. The theoretical model considers a seven-level atomic system of $^{85}$Rb confined in the thin cell (see Fig. 1).

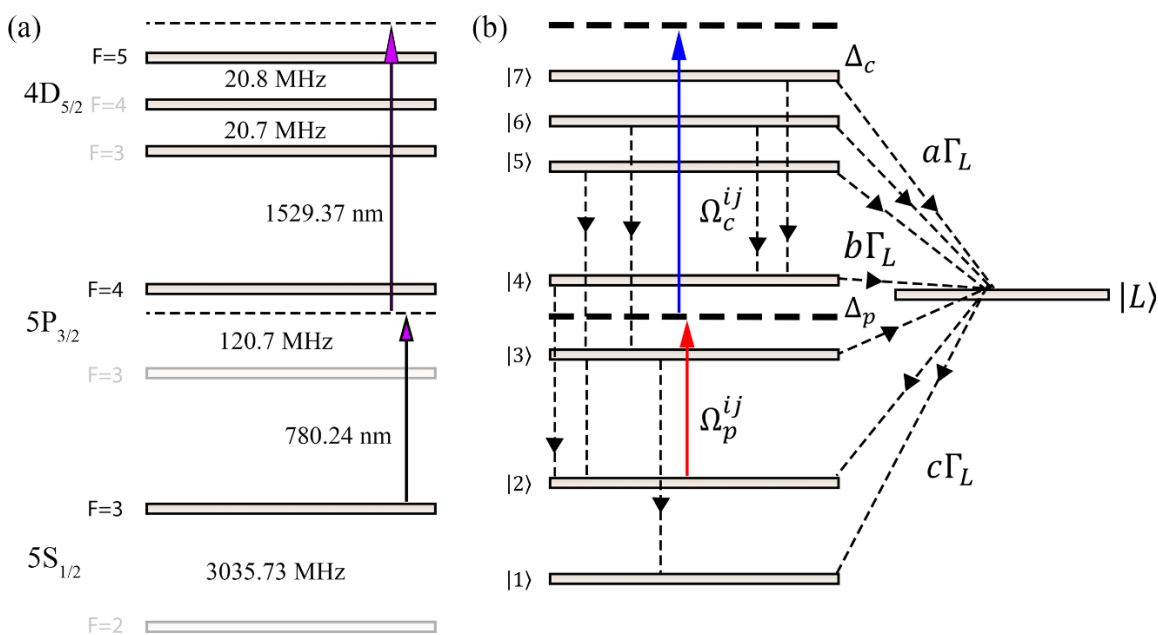

**FIG. 2. Experimental and theoretical energy-level schemes of the near-infrared rubidium system** (a) Experimental energy-level scheme of $^{85}$Rb. The probe beam at 780.24 nm is tuned slightly (<1 MHz) below the $|5S_{1/2}, F = 3\rangle \rightarrow |5P_{3/2}, F = 4\rangle$ transition, while the coupling beam at 1529.37 nm is scanned across the $|4D_{5/2}, F = 3,4,5\rangle$ hyperfine manifold. (b) Energy-level configuration used in the theoretical model. $\Omega_p$ and $\Omega_c$ denote the Rabi frequencies of the probe and coupling fields, respectively, and $\Delta_p$ and $\Delta_c$ are their corresponding detunings. The coefficients (a), (b), and (c) represent the branching ratios from the five upper states $|3\rangle - |7\rangle$ to the ground states and to the loss state $|L\rangle$.

The atomic medium interacts with two optical fields: a probe beam that is slightly red-detuned ($\Delta_p < 1$ MHz) from the $|5S_{1/2}, F = 3\rangle \rightarrow |5P_{3/2}, F = 4\rangle$ intermediates states transition, and a coupling beam that is scanned across the $|5P_{3/2}, F = 4\rangle \rightarrow |4D_{5/2}, F = 3,4,5\rangle$ manifold. Within the density-matrix formalism, and under the dipole and rotating-wave approximations, the time evolution of the atomic populations and coherences is governed by the quantum master equation

(1) $$\frac{\partial \rho}{\partial t} = -\frac{i}{\hbar}[\mathcal{H}, \rho] + \mathcal{L}\rho$$

Here $\mathcal{H}$ is the atomic Hamiltonian and can be describe as

(2) $$\mathcal{H} = -\omega_{HFS}|1\rangle\langle 1| + (\Delta_p - \delta_1)|3\rangle\langle 3| + \Delta_p|4\rangle\langle 4|$$
$$+ \left(\Delta_p + \Delta_c + (\delta_2 + \delta_3)\right)|5\rangle\langle 5| + \left(\Delta_p + \Delta_c + \delta_3\right)|6\rangle\langle 6|$$
$$+ (\Delta_p + \Delta_c)|7\rangle\langle 7| + \Omega_p^{24}|2\rangle\langle 4| + \Omega_c^{45}|4\rangle\langle 5| + \Omega_c^{46}|4\rangle\langle 6|$$
$$+ \Omega_c^{47}|4\rangle\langle 7| + \text{H.c}$$

Where $\omega_{HFS} = 2\pi \cdot 3.035$ GHz for $^{85}$Rb, $\Omega_c^{ij}$ is the coupling Rabi frequencies define as $\Omega_c^{ij} = d_{ij} \cdot E_c/\hbar$ and $\Omega_p^{ij} = d_{ij} \cdot E_P/\hbar$ are the Rabi frequencies of the probe beam and the pump beam respectively. $d_{ij}$ is the moment dipole of the transition and it is equal to

(3) $$d_{ij} = \langle F_i|e \cdot r|F_j\rangle = \langle J_i|e \cdot r|J_j\rangle(-1)^{F_j+J_i+1+I} \cdot \left(S_{F_iF_j}\right)^{1/2}$$

Here $I, F, J$ are the angular momentum quantum numbers, $S_{FF'}$ is transition strength factor (for the transition strength calculations see the appendix), the reduced matrix elements $\langle J_i|e \cdot r|J_j\rangle$ is equal to[37]:

(4) $$\langle J_i|e \cdot r|J_j\rangle = \frac{3\pi\varepsilon_0\hbar c^3}{\omega_0^3 \tau}\frac{2J_j + 1}{2J_i + 1}$$

Here $\omega_0$ is resonance transition frequency, $\tau$ is the natural lifetime. $\Delta_p, \Delta_c$ is the detuning of the probe and the coupling beam respectively. $\delta_i$ are the level spacing between two adjacent energy levels (see Fig. 2). H.c is the Hermitian conjugate. $\mathcal{L}\rho$ is Lindblad operator (dissipative processes) which describes population and coherence time decay. It is equal to

(5) $$\mathcal{L}\rho = \sum_d \frac{\Gamma_{tot}}{2}\left(\sigma_d^\dagger \sigma_d \rho + \rho \sigma_d^\dagger \sigma_d - 2\sigma_d \rho \sigma_d^\dagger\right)$$

where $\sigma_d^\dagger = |n\rangle\langle m|$ and $\Gamma_{tot}$ is the total decaying rates from state $n, m \in \{|1\rangle, |2\rangle, |3\rangle, |4\rangle, |5\rangle, |6\rangle, |7\rangle\}$. The decaying rates can be described as

$$\Gamma_{tot} = \Gamma_0 + \Gamma_{self} + \Gamma_L \tag{6}$$

Where $\Gamma_0$ is the spontaneous emission of the excited states ($\Gamma_{5P_{3/2}} = 2\pi \cdot 6.06$ MHz and $\Gamma_{4D_{5/2}} = 2\pi \cdot 1.97$ MHz), $\Gamma_{self}$ is the self-broadening relaxation rate. Since our all of our experiments done under 120C, as shown by Lee Weller et. al [38] the self-broadening relaxation rate is equal to $\Gamma_{self} \approx 2\pi \cdot 1$ MHz, hence, we can neglect the influence of self-broadening. Finally, the dominant relaxation rate is due to wall collision relaxation rate and can be described as

$$\Gamma_L = 2\pi \cdot \frac{v_z}{L_z/2} \tag{7}$$

Here $L_z$ is the cell thickness, and $v_z$ is the mean atomic velocity along the z direction. At temperature of 120C the mean atomic velocity along the z direction is around $v_z \approx 277$ [m/s].

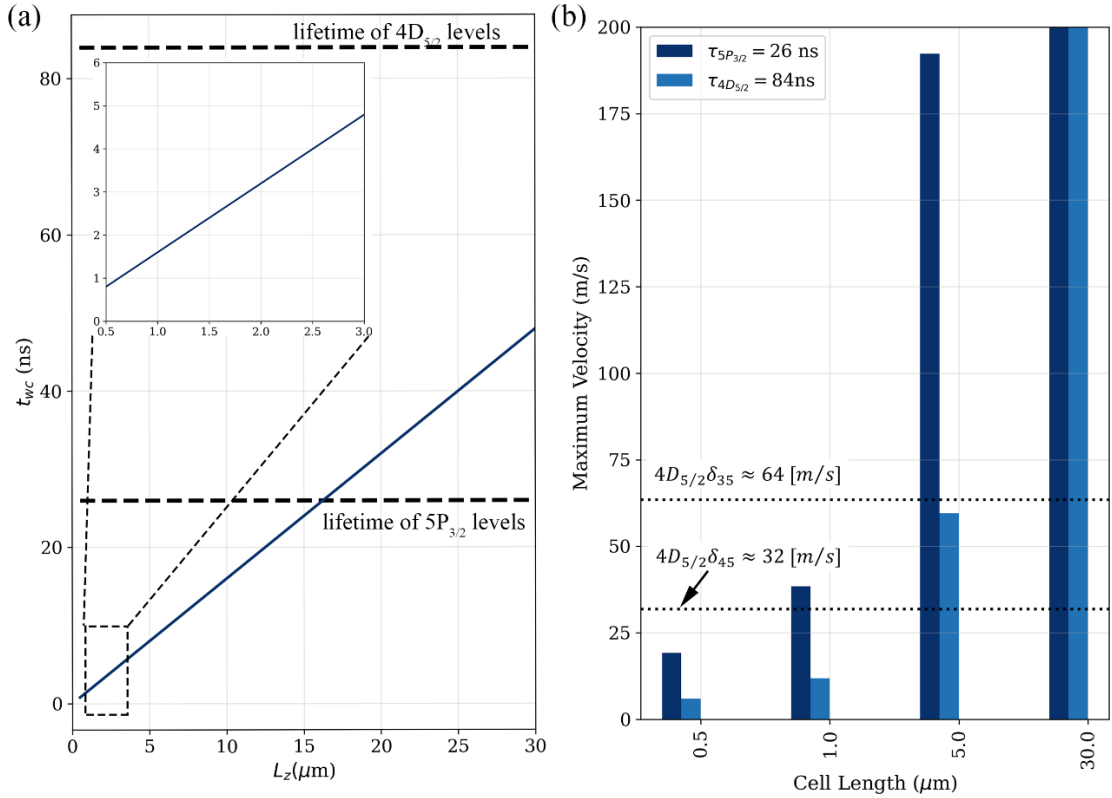

**FIG. 3. Wall-collision–induced relaxation and velocity filtering in ultrathin vapor cells** (a) Calculated wall-collision relaxation rate at a temperature of 120°C ($v_z \approx 277$ m/s) as a function of cell thickness. The black dashed lines indicate the natural lifetimes of the intermediate $|5P_{3/2}\rangle$ state and the upper excited

$|4D_{5/2}\rangle$ state. (b) Maximum allowed longitudinal atomic velocity $v_z$ for atoms moving perpendicular to the cell windows, determined by the condition $v_{max} = L_z/\tau_{4D_{5/2}}$ for the upper excited state and $v_{max} = L_z/\tau_{5P_{3/2}}$ for the intermediate state. The black dashed lines denote the energy separations between the hyperfine manifolds of the $|4D_{5/2}\rangle$ states.

As shown in Fig. 3(a), at a temperature of 120 °C the wall-collision time for cell thicknesses ranging from 500 nm to 15 μm is shorter than the spontaneous lifetime of the $5P_{3/2}$ excited state. Consequently, optical-pumping mechanisms that normally deplete the population of the cycling transition $|5S_{1/2}, F = 3\rangle \rightarrow |5P_{3/2}, F = 4\rangle$ via decay through non-cycling intermediate states into the uncoupled ground state $|5S_{1/2}, F = 2\rangle$ are strongly suppressed. As a result, the population of the closed (cycling) transition remains essentially unchanged. Moreover, even when the probe laser is tuned to the cycling transition $|5S_{1/2}, F = 3\rangle \rightarrow |5P_{3/2}, F = 4\rangle$, non-cycling excitation can occur for atoms whose Doppler shift matches the hyperfine splitting between the $|5P_{3/2}, F = 4\rangle$ and $|5P_{3/2}, F = 3\rangle$ levels, which is 120.7 MHz. This condition is given by

$$\omega_{F=3 \rightarrow F=4} = -\mathbf{k}_p \cdot \mathbf{v} \Rightarrow v = \frac{\omega_{F=3 \rightarrow F=4}}{k_p} \quad (8)$$

where $k_p$ is the probe-beam wave vector, $k_p = 2\pi/\lambda_p = 2\pi/(780 \times 10^{-9}\text{ m})$. Substituting these values into Eq. (4) yields a minimum atomic velocity of $v \approx 94$ m/s. As indicated in Fig. 3(a), in ultrathin cells such velocities are effectively filtered out by rapid wall collisions. Consequently, non-cycling transitions are strongly suppressed, and only the cycling states remain significantly populated. Fig. 3(b) shows the maximum atomic velocity along the z direction that contributes to the intermediate and excited-state populations as a function of cell thickness. For a cell thickness of 5 μm, the maximum allowed velocity is approximately 60 m/s. At this velocity, Doppler shifts are sufficient to enable excitation of atoms from the $4D_{5/2}, F = 4$ hyperfine manifold by the coupling laser, allowing population of non-cycling states. This Doppler-induced coupling degrades the formation of an effective two-level system (TLS). In contrast, for cell thicknesses below 5 μm, the maximum allowed velocity is reduced to approximately 12 m/s. Under these conditions, the Doppler shift is too small to bridge the hyperfine splitting, and non-cycling transitions are therefore strongly suppressed. As a result, only the cycling transition remains active, enabling the robust realization of an effective TLS. The calculated spectra of the DROP, FDROP, and fluorescence signals are obtained by solving the full set of master equations, whose derivation is provided in the Appendix. Our measurements rely on detecting either the optical transmission or the fluorescence emitted from ultrathin vapor cells. Accordingly, the theoretical description requires solving the density-matrix equations under steady-state conditions ($\partial \rho/\partial t = 0$), To analyze the response of the thin cell under the interaction of two optical fields, we calculate three observables: (i) the DROP signal of the probe beam, (ii) the fluorescence induced by the probe beam (FDROP), and (iii) the fluorescence generated by the coupling beam at 1529 nm. The DROP signal is proportional to the absorption of the probe beam and is expressed as

$$\text{DROP} = \int_0^\infty Im[\rho_{24}(v)]dz \int_{-\infty}^\infty W(v)\,dv \quad (9)$$

where $W(v)$ is the Boltzmann velocity distribution from our thin cell. The fluorescence emitted from the $4D_{5/2}$ manifold is given by

$$\text{Fluorescence} = \sum_{i=5}^{7} \int_0^\infty \rho_{ii} dz \int_{-\infty}^\infty W(v)\,dv \quad (10)$$

Where we are summing over the population density matrix elements ($\rho_{ii}, i \in 5,6,7$). Finally, the FDROP signal, corresponding to the fluorescence induced by the probe beam, is calculated as

$$\text{FDROP} = \sum_{i=3}^{5} \int_0^\infty \rho_{ii} dz \int_{-\infty}^\infty W(v)dv \quad (11)$$

where the summation includes the relevant intermediate and excited states. The Boltzmann velocity distribution is equals $W(v) = (1/u\sqrt{\pi})\exp(-v^2/u^2)$ here u is the most probable velocity $u = \sqrt{2k_B T/m}$ where m is the mass of the atoms and T their absolute temperature.

**Experimental setup and results**

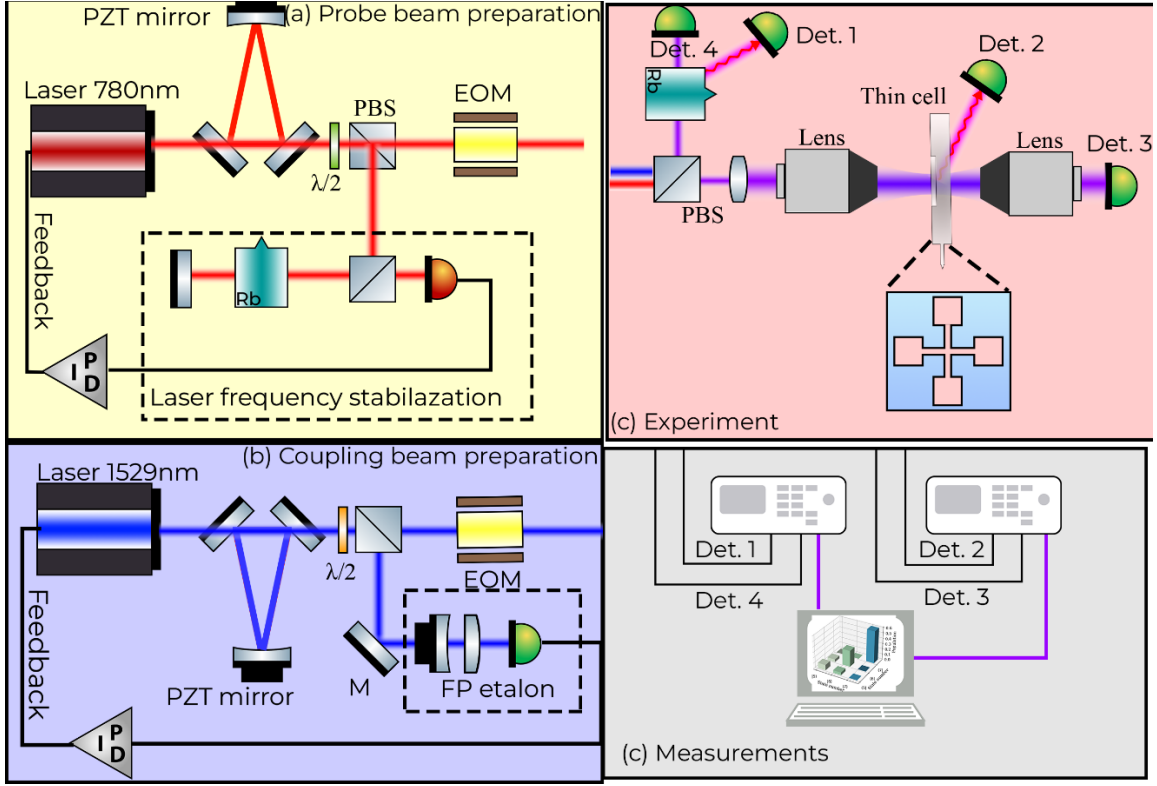

**FIG. 4. Experimental setup for OODR, DROP, and FDROP spectroscopy in ultrathin rubidium vapor cells** Experimental setup for OODR, DROP, and FDROP spectroscopy of Rb atoms. (a) Probe-beam preparation: the probe laser is frequency-stabilized near the cycling transition $|5S_{1/2}, F = 3\rangle \to |5P_{3/2}, F = 4\rangle$ using saturated-absorption spectroscopy and then sent through an electro-optic modulator (EOM) for intensity stabilization and frequency scanning around resonance. (b) Coupling-beam preparation: the coupling laser is stabilized to the excited-state cycling transition using a Fabry–Perot (FP) etalon together with a resonance signal from a reference vapor cell. (c) Interaction region: the probe and coupling beams counter-propagate through either an ultrathin vapor cell or a reference cell; both cells are placed inside magnetic shielding to suppress external magnetic-field effects. (d) Detection: transmitted and fluorescence signals are recorded with photodetectors and an oscilloscope and analyzed on a computer. EOM: electro-optic modulator; M: mirror; FP: Fabry–Perot etalon; PBS: polarizing beam splitter; Det: detector; λ/2 plate: half-wave plate.

The experimental setup is shown schematically in Fig. 4 and consists of the probe-beam preparation, coupling-beam preparation, and the interaction and detection stages. The probe beam, shown in Fig. 4(a), is tuned near the $|5S_{1/2}, F = 3\rangle \to |5P_{3/2}, F = 4\rangle$ transition of natural Rb and is generated by a 780 nm external-cavity diode laser (ECDL) with a typical linewidth of ~200 kHz providing up to 100 mW of optical power. To suppress beam-pointing noise and higher-order transverse modes, the probe beam is coupled into a high-finesse three-mirror ring cavity acting as a spatial mode cleaner, which selectively enhances the fundamental $LG_{00}$ mode while rejecting higher-order modes. A small fraction of the probe light is diverted to a saturated-absorption spectroscopy setup using a rubidium reference cell, allowing stabilization of the laser frequency close to the desired atomic resonance. The main probe beam then passes through an electro-optic modulator (EOM), which is used for phase modulation and active stabilization of the probe-beam power before the beam is delivered to the interaction region. The coupling beam at 1529 nm, illustrated in Fig. 4(b), is generated using a continuously tunable laser (CTL) with a typical linewidth of ~100 kHz . As in the probe-beam path, the coupling beam is spatially filtered using a three-mirror ring cavity to ensure a clean fundamental transverse mode. Frequency stabilization and monitoring of the coupling laser are achieved using a Fabry–Perot etalon with a free spectral range of 200 MHz. An electro-optic modulator is also employed in the coupling-beam path, serving the same dual role of optical modulation and power stabilization prior to the experiment. As shown in Fig. 4(c), both the probe (20 μW) and coupling beams (60 μW) are split into two paths. One path is directed to a 7.5 cm-long reference cell containing natural Rb vapor placed inside a magnetic shielding to suppressed stray magnetic field, which is used for auxiliary spectroscopy and system diagnostics. The second path guides the probe and coupling beams into the main experimental region, where they are mode-matched and made to co-propagate using a set of lenses. The beams interact in a homemade ultrathin vapor cell containing natural rubidium vapor also placed in magnetic shielding and no buffer gas, enabling strong confinement along the propagation direction and access to velocity-selective and surface-dominated interaction regimes. Finally, the transmitted and fluorescence signals from both the reference and ultrathin cells are detected using photodetectors positioned as indicated in Fig. 4(c). All detector signals are recorded using a digital oscilloscope and analyzed on a computer, as shown in Fig. 4(d), allowing simultaneous acquisition and comparison of transmission and fluorescence signals from the different measurement channels.

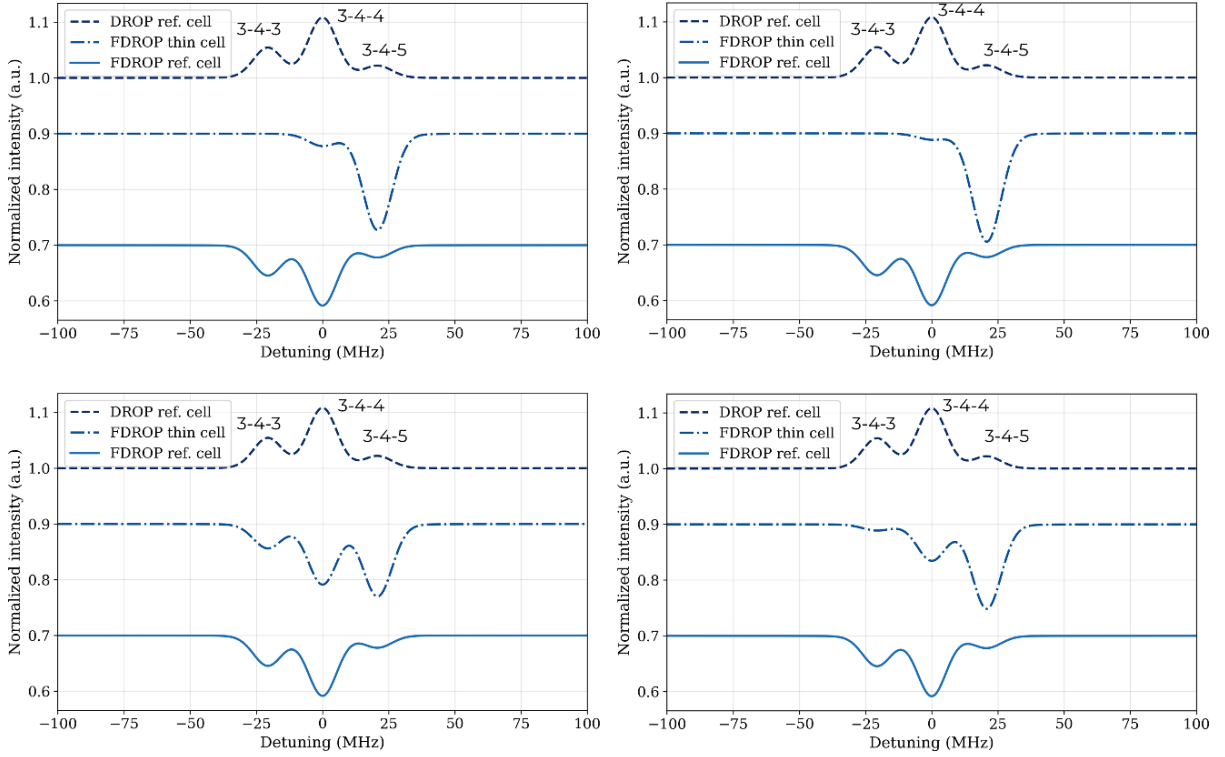

**FIG. 5. Thickness-dependent theoretical FDROP spectra and comparison with a reference cell**
Theoretical FDROP spectra calculated for different vapor-cell thicknesses, compared with DROP and FDROP spectra obtained from a reference cell. The simulations are performed using probe and coupling Rabi frequencies $\Omega_p = 2\pi \times 4$ MHz and $\Omega_c = 2\pi \times 20$ MHz, respectively. The relaxation and decay parameters are $\Gamma_{self} \approx 2\pi \times 1$ MHz, $\Gamma_{5P_{3/2}} = 2\pi \cdot 6.06$ MHz and $\Gamma_{4D_{5/2}} = 2\pi \cdot 1.97$ MHz. Wall-collision–induced relaxation is included through phenomenological factors (a = b = c = 0.5).

Fig. 5 presents the theoretical FDROP spectra obtained from our homemade ultrathin vapor cell for thicknesses ranging from 500 nm to 30 μm and compares them with the corresponding DROP and FDROP spectra measured in a reference cell (7.5 cm). Although the cycling transition $|5P_{3/2}, F = 4\rangle \rightarrow |4D_{5/2}, F = 5\rangle$ has the largest dipole transition strength (see Appendix), the simulations reveal that, in the reference cell, the dominant contribution arises instead from the non-cycling transition $|5S_{1/2}, F = 3\rangle \rightarrow |5P_{3/2}, F = 4\rangle \rightarrow |4D_{5/2}, F = 4\rangle$. This behavior originates from two main mechanisms. First, optical pumping from the intermediate state transfers population to the uncoupled ground level $|5S_{1/2}, F = 2\rangle$, leading to population depletion of the cycling channel and an enhanced contribution of the intermediate-state transitions, which manifests as a stronger DROP signal. Second, the small hyperfine splitting of the excited

$4D_{5/2}$ manifold, combined with the large Doppler broadening in the macroscopic reference cell, allows atoms from different velocity classes to simultaneously satisfy the resonance condition for the non-cycling transition. This Doppler-induced overlap increases absorption in the non-cycling channel and further suppresses the effective strength of the cycling transition. As a result, despite its intrinsically larger transition strength, the cycling transition is strongly reduced in the reference cell, whereas the non-cycling transition dominates the observed FDROP spectra. In contrast, when operating in the ultrathin-cell regime, as shown in Fig. 3, the FDROP signal originates almost exclusively from slow atoms whose longitudinal velocities are strongly restricted by frequent wall collisions. Consequently, once the laser frequency is tuned to the cycling transition, absorption arising from Doppler-shifted velocity classes associated with off-resonant (non-cycling) transitions is strongly suppressed. Under these conditions, the intrinsic transition strength becomes the dominant factor determining the observed signal. Moreover, the reduced interaction time in ultrathin cells significantly suppresses optical pumping of the intermediate state into uncoupled ground levels. As a result, population accumulation in non-active states is strongly inhibited, further favoring the cycling transition. Together, these effects restore the dominance of the cycling transition in the FDROP spectra, in sharp contrast to the behavior observed in macroscopic reference cells.

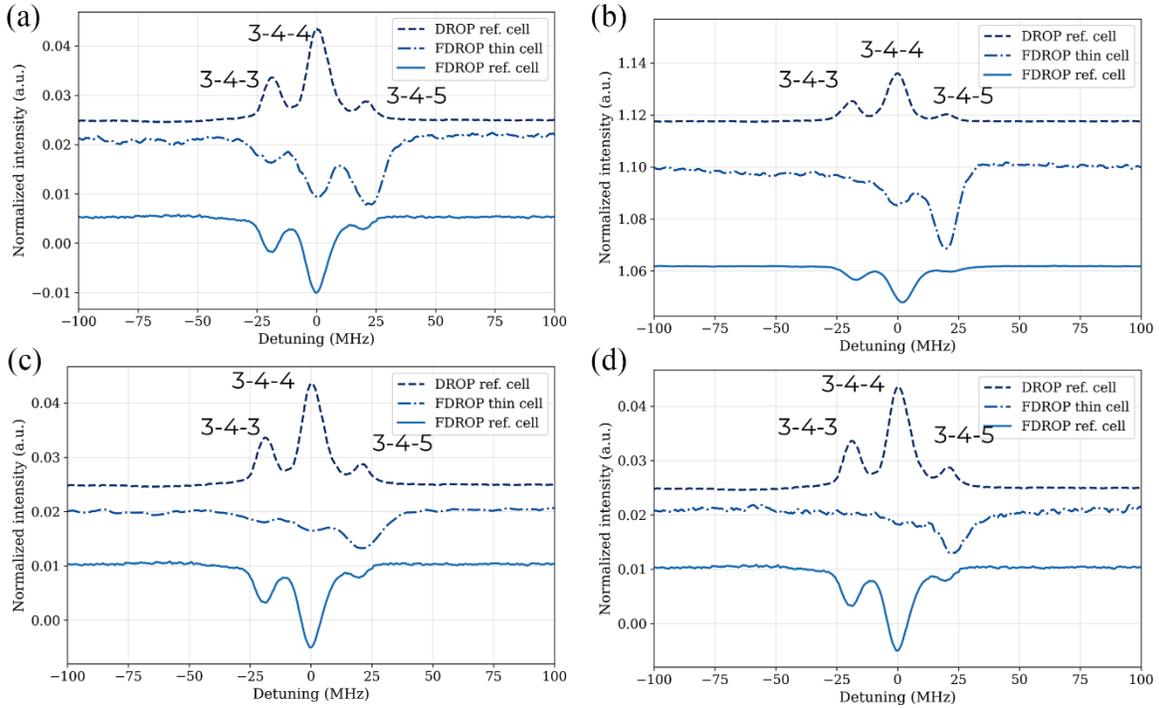

**FIG. 6. Experimental FDROP spectra as a function of vapor-cell thickness**. Experimental FDROP spectra measured for different vapor-cell thicknesses, compared with the corresponding DROP and FDROP spectra obtained from a reference cell.

Figure 6 presents the experimental results obtained from our ultrathin vapor cells together with measurements from a reference cell, showing good agreement with the theoretical model. The ultrathin cells were operated at a temperature of 120 °C, corresponding to an atomic density of approximately $1 \times 10^{13}$ cm$^{-3}$, under which the self-broadening effect can be safely neglected[38]. For these measurements, both the probe and coupling beam powers were set to 500 μW. In contrast, the reference cell was maintained at room temperature (25 °C) while using the same probe and coupling beam powers as in the ultrathin-cell measurements.

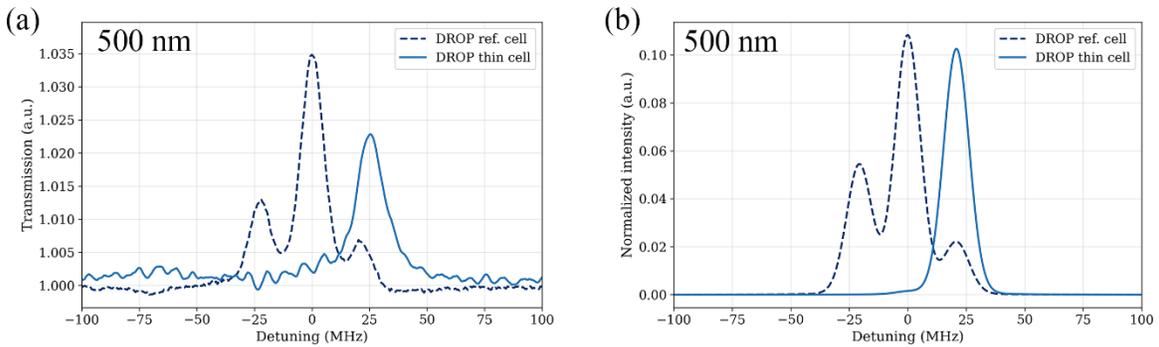

**FIG. 7.** (a) Experimental and (b) theoretical DROP spectra obtained from a 500-nm-thick vapor cell, compared with the corresponding DROP spectra measured from a reference cell.

Figure 7 presents the experimental and theoretical DROP spectra obtained from the reference cell and from the 500 nm thick cell. In both the measured and simulated spectra, the dominant feature corresponds to the cycling transition $|5P_{3/2}, F = 4\rangle \rightarrow |4D_{5/2}, F = 5\rangle$, as expected from its larger transition strength. The results clearly indicate that, in the ultrathin cell, frequent atom–wall collisions strongly suppress optical pumping into the uncoupled ground state $|5S_{1/2}, F = 2\rangle$. In addition, excitation of nearby non-cycling transitions is inhibited, since only slow atoms with small Doppler shifts contribute significantly to the signal.

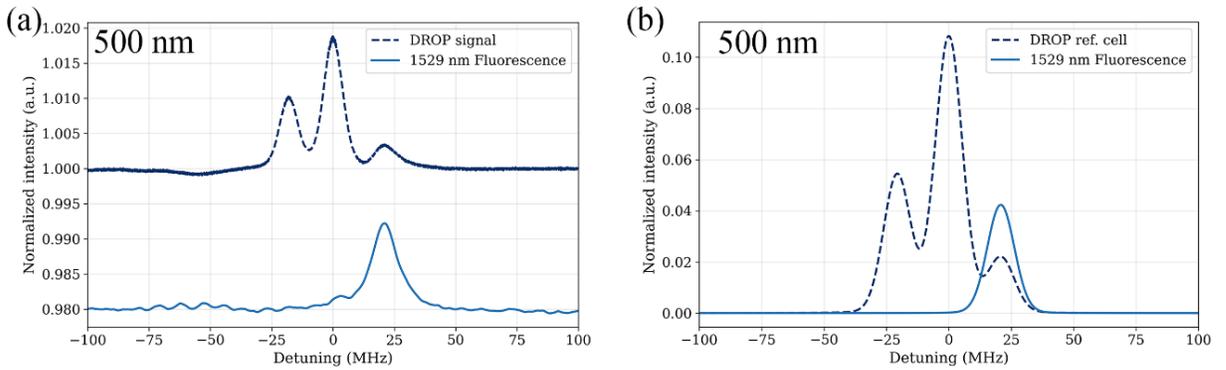

**FIG. 8. Experimental and theoretical DROP spectra for a 500-nm-thick vapor cell** (a) Experimental and (b) theoretical OODR fluorescence spectra obtained from a 500-nm-thick vapor cell, compared with the corresponding DROP spectra measured from a reference cell.

Fig. 8 presents the experimental and theoretical OODR fluorescence spectra at 1529 nm obtained from the 500 nm thick vapor cell. A clear and strong agreement between the measured data and the theoretical model is observed. The fluorescence signal is dominated by the cycling transition, indicating that this transition provides the primary contribution to the detected emission. In the ultrathin-cell regime, frequent atom wall collisions efficiently suppress optical pumping into uncoupled ground states, while the strong velocity-selective filtering favors slow atoms with minimal Doppler shifts. As a result, population transfer through non-cycling or nearby transitions is strongly reduced, and the fluorescence originates predominantly from the closed cycling transition. This behavior highlights the crucial role of confinement and wall-induced relaxation in enhancing spectral selectivity an simplifying the effective atomic dynamics in ultrathin vapor cells.

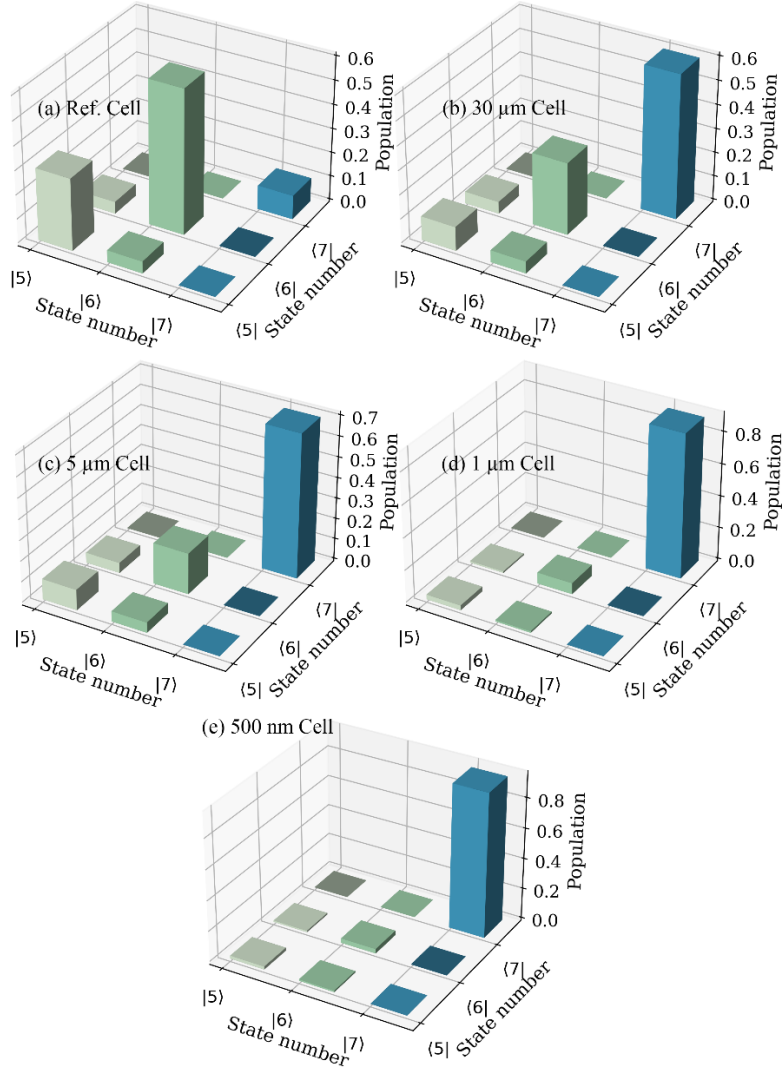

**FIG. 9. Density-matrix reconstruction as a function of vapor-cell thickness**. Density-matrix reconstruction for different vapor-cell thicknesses, obtained by combining information from DROP and FDROP measurements together with fluorescence detected from the excited states.

Finally, as shown in Fig. 9, we reconstruct the atomic density matrix by combining all the previous experimental and theoretical results. Figure 9 clearly demonstrates that the population of state $|7\rangle$, corresponding to the $|4D_{5/2}, F = 5\rangle$ level, increases systematically as the cell thickness is reduced. In particular, for a cell thickness of 500 nm, state $|7\rangle$ becomes the dominant populated level, indicating that the dynamics are governed primarily by the cycling transition. In contrast, for the reference cell, the largest population is found in state $|6\rangle$, which corresponds to the $|4D_{5/2}, F = 4\rangle$ transition. This comparison highlights the profound impact of spatial confinement on the population redistribution among excited states, where enhanced wall-collision-induced relaxation and velocity selectivity in ultrathin cells favor the cycling transition over competing non-cycling channels.

**Conclusion**

In this work, we have presented a comprehensive experimental and theoretical study of DROP and FDROP spectroscopy in rubidium vapor confined to ultrathin cells with thicknesses down to 500 nm and compared the results with those obtained from a conventional macroscopic reference cell. By combining transmission, DROP, FDROP, and fluorescence measurements with a density-matrix model, we have shown how extreme spatial confinement fundamentally modifies the atomic dynamics and the resulting spectral response. In the ultrathin-cell regime, frequent atom–wall collisions and the associated relaxation processes give rise to strong velocity-selective filtering, such that only slow atoms with small Doppler shifts contribute significantly to the signal. As a result, effective Doppler broadening is strongly suppressed, optical pumping into uncoupled ground states is minimized, and population leakage to nearby non-cycling transitions is largely inhibited.

Under these conditions, the spectroscopy becomes dominated by the closed cycling transition $|5P_{3/2}, F = 4\rangle \rightarrow |4D_{5/2}, F = 5\rangle$, in stark contrast to the behavior observed in macroscopic reference cells, where Doppler broadening and optical pumping favor non-cycling pathways. The DROP and FDROP spectra, together with the 1529 nm fluorescence measurements, show excellent agreement with our theoretical model across all the cell thicknesses investigated. Reconstruction of the atomic density matrix further reveals a clear redistribution of excited-state populations with decreasing cell thickness, culminating in a dominant population of the $|4D_{5/2}, F = 5\rangle$ level for the 500 nm cell, while the reference cell remains dominated by the $|4D_{5/2}, F = 4\rangle$ level.

These observations provide direct evidence of a transition from Doppler-dominated dynamics in macroscopic vapor cells to collision- and confinement-dominated dynamics in ultrathin cells. In contrast to conventional excited-state spectroscopy in macroscopic vapor cells, where Doppler broadening and optical pumping harm cycling transitions, our results demonstrate that ultrathin confinement reverses this hierarchy and restores a closed, effective two-level atomic response at telecom wavelengths. Overall, this work establishes DROP and FDROP spectroscopy as powerful tools for probing and controlling atomic populations in strongly confined thermal vapors. The demonstrated suppression of Doppler and optical-pumping effects, together with enhanced selectivity toward cycling transitions, opens new opportunities for realizing simplified effective two-level systems, compact near-infrared light sources, and precision quantum-optical platforms based on ultrathin alkali-vapor cells.

# Supplementary information

# Engineering Near-Infrared Two-Level Systems in Confined Alkali Vapors

Gilad Orr[1], Golan Ben-Ari[2], Eliran Talker[2]

[1]Department of Physics, Ariel University, Ariel 40700

[2]Department of Electrical Engineering, Ariel University, Ariel 40700, Israel

*Corresponding author: elirant@ariel.ac.il

## S1) Lindblad master equation

the master equation of our seven-level system can be described as

$$\frac{\partial \rho}{\partial t} = -\frac{i}{\hbar}[\mathcal{H}, \rho] + G\rho$$

Here $\mathcal{H} = \mathcal{H}_0 + V$ where $\mathcal{H}_0$ is the unperpetrated Hamiltonian and is equal

$$\mathcal{H}_0 = \begin{pmatrix} -\omega_{HFS} & 0 & 0 & 0 & 0 & 0 & 0 \\ 0 & 0 & 0 & 0 & 0 & 0 & 0 \\ 0 & 0 & \Delta_p - \delta_1 & 0 & 0 & 0 & 0 \\ 0 & 0 & 0 & \Delta_p & 0 & 0 & 0 \\ 0 & 0 & 0 & 0 & \Delta_p + \Delta_c + (\delta_2 + \delta_3) & 0 & 0 \\ 0 & 0 & 0 & 0 & 0 & \Delta_p + \Delta_c + \delta_3 & 0 \\ 0 & 0 & 0 & 0 & 0 & 0 & \Delta_p + \Delta_c \end{pmatrix}$$

$\Delta_p, \Delta_c$ is the detuning of the probe and the conjugate beams, $\delta_i$ is the level spacing between the adjacent levels The interaction potential can be described as

$$V = \begin{pmatrix} 0 & 0 & 0 & 0 & 0 & 0 & 0 \\ 0 & 0 & 0 & \Omega_p & 0 & 0 & 0 \\ 0 & 0 & 0 & 0 & 0 & 0 & 0 \\ 0 & \Omega_p & 0 & 0 & \Omega_c^{45} & \Omega_c^{46} & \Omega_c^{47} \\ 0 & 0 & 0 & \Omega_c^{45} & 0 & 0 & 0 \\ 0 & 0 & 0 & \Omega_c^{46} & 0 & 0 & 0 \\ 0 & 0 & 0 & \Omega_c^{47} & 0 & 0 & 0 \end{pmatrix}$$

Where $\Omega_c^{ij} = \left(\Omega_c^{ij}\right)^*$ and $\Omega_p = \Omega_p^*$. The relaxation matrix, G, in the Liouville equation given by:

$$G = \begin{pmatrix} \Gamma_{12} & 0 & 0 & 0 & 0 & 0 & 0 \\ 0 & \Gamma_{12} & 0 & 0 & 0 & 0 & 0 \\ 0 & 0 & \gamma_3 + b\Gamma_L & 0 & 0 & 0 & 0 \\ 0 & 0 & 0 & \gamma_4 + b\Gamma_L & 0 & 0 & 0 \\ 0 & 0 & 0 & 0 & \gamma_5 + c\Gamma_L & 0 & 0 \\ 0 & 0 & 0 & 0 & 0 & \gamma_6 + c\Gamma_L & 0 \\ 0 & 0 & 0 & 0 & 0 & 0 & \gamma_7 + c\Gamma_L \end{pmatrix}$$

The equations of motion for the population and coherence terms are

$$\dot\rho_{11} = -\Gamma_{12}\rho_{11} + \gamma_3\rho_{33} + \Gamma_{12}\rho_{22}$$

$$\dot\rho_{12} = -\Gamma_{12}\rho_{12} - i(\omega_{HFS}\rho_{12} + \Omega_p\rho_{14})$$

$$\dot\rho_{13} = -\frac{1}{2}\rho_{13} \cdot \left(\Gamma_{12} + \gamma_3 - 2i(\delta 1 - \Delta_p - \omega_{HFS})\right)$$

$$\dot\rho_{14} = \frac{1}{2}\Big(-\Gamma_{12}\rho_{14} - \gamma_4\rho_{14} \\ - 2i\big(\Delta_p\rho_{14} + \rho_{15}\Omega_c^{45} + \rho_{16}\Omega_c^{46} + \rho17\Omega_c^{47} + \rho14\omega_{HFS} + \rho_{12}\Omega_p\big)\Big)$$

$$\dot\rho_{15} = \frac{1}{2}\Big(-\Gamma_{12}\rho_{15} - \gamma_5\rho_{15} \\ - 2i\big(\delta_2\rho_{15} + \delta_3\rho_{15} + \Delta_c\rho_{15} + \Delta_p\rho_{15} + \rho_{14}\Omega_c^{45} + \rho_{15}\omega_{HFS}\big)\Big)$$

$$\dot\rho_{16} = \frac{1}{2}\Big(-\Gamma_{12}\rho_{16} - \gamma_6\rho_{16} - 2i\big(\delta_3\rho_{16} + \Delta_c\rho_{16} + \Delta_p\rho_{16} + \rho_{14}\Omega_c^{46} + \rho_{16}\omega_{HFS}\big)\Big)$$

$$\dot\rho_{17} = \frac{1}{2}\Big(-\Gamma_{12}\rho_{17} - \gamma_7\rho_{17} - 2i\big(\Delta_c\rho_{17} + \Delta_p\rho_{17} + \rho_{14}\Omega_c^{47} + \rho_{17}\omega_{HFS}\big)\Big)$$

$$\dot\rho_{21} = -\Gamma_{12}\rho_{21} + i(\rho_{21}\omega_{HFS} + \rho_{41}\Omega_p)$$

$$\dot\rho_{22} = -\Gamma_{12}\rho_{22} - i(\rho_{24} - \rho_{42})\Omega_p + \gamma_3\rho_{33} + \gamma_4\rho_{44} + \Gamma_{12}\rho_{11}$$

$$\dot\rho_{23} = -\frac{1}{2}(\Gamma_{12} + \gamma_3)\rho_{23} + i(\delta_1 - \Delta_p)\rho_{23} + i\rho_{43}\Omega_p$$

$$\dot\rho_{24} = \frac{1}{2}\Big(-(\Gamma_{12} + \gamma_4 + 2i\Delta_p)\rho_{24} - 2i\big(\rho_{25}\Omega_c^{45} + \rho_{26}\Omega_c^{46} + \rho_{27}\Omega_c^{47} + \Omega_p(\rho_{22} - \rho_{44})\big)\Big)$$

$$\dot\rho_{25} = -\frac{1}{2}(\Gamma_{12} + \gamma_5)\rho_{25} - i(\delta_2 + \delta_3 + \Delta_c + \Delta_p)\rho_{25} - i\rho_{24}\Omega_c^{45} + i\rho_{45}\Omega_p$$

$$\dot\rho_{26} = -\frac{1}{2}(\Gamma_{12} + \gamma_6)\rho_{26} - i(\delta_3 + \Delta_c + \Delta_p)\rho_{26} - i\rho_{24}\Omega_c^{46} + i\rho_{46}\Omega_p$$

$$\dot\rho_{27} = \frac{1}{2}\Big(-\Gamma_{12}\rho_{27} - \gamma_7\rho_{27} - 2i\big(\Delta_c\rho_{27} + \Delta_p\rho_{27} + \rho_{24}\Omega_c^{47} - \rho_{47}\Omega_p\big)\Big)$$

$$\dot\rho_{31} = -\frac{1}{2}\rho_{31}\left(\Gamma_{12} + \gamma_3 + 2i(\delta_1 - \Delta_p - \omega_{\text{HFS}})\right)$$

$$\dot\rho_{32} = -\frac{1}{2}(\Gamma_{12} + \gamma_3)\rho_{32} + i(-\delta_1 + \Delta_p)\rho_{32} - i\rho_{34}\Omega_p$$

$$\dot\rho_{33} = -\gamma_3\rho_{33} + \gamma_5\rho_{55} + \gamma_6\rho_{66}$$

$$\dot\rho_{34} = \frac{1}{2}\left(-\gamma_3\rho_{34} - \gamma_4\rho_{34} - 2i\left(\delta_1\rho_{34} + \rho_{35}\Omega_c^{45} + \rho_{36}\Omega_c^{46} + \rho_{37}\Omega_c^{47} + \rho_{32}\Omega_p\right)\right)$$

$$\dot\rho_{35} = -\frac{1}{2}(\gamma_3 + \gamma_5)\rho_{35} - i(\delta_1 + \delta_2 + \delta_3 + \Delta_c)\rho_{35} - i\rho_{34}\Omega_c^{45}$$

$$\dot\rho_{36} = -\frac{1}{2}(\gamma_3 + \gamma_6)\rho_{36} - i(\delta_1 + \delta_3 + \Delta_c)\rho_{36} - i\rho_{34}\Omega_c^{46}$$

$$\dot\rho_{37} = \frac{1}{2}\left(-\gamma_3\rho_{37} - \gamma_7\rho_{37} - 2i\left(\delta_1\rho_{37} + \Delta_c\rho_{37} + \rho_{34}\Omega_c^{47}\right)\right)$$

$$\dot\rho_{41} = \frac{1}{2}\Big(-\Gamma_{12}\rho_{41} - \gamma_4\rho_{41} \\ + 2i\big(\Delta_p\rho_{41} + \rho_{51}\Omega_c^{45} + \rho_{61}\Omega_c^{46} + \rho_{71}\Omega_c^{47} + \rho_{41}\omega_{\text{HFS}} + \rho_{21}\Omega_p\big)\Big)$$

$$\dot\rho_{42} = \frac{1}{2}\Big(-\Gamma_{12}\rho_{42} - \gamma_4\rho_{42} \\ + 2i\big(\Delta_p\rho_{42} + \rho_{52}\Omega_c^{45} + \rho_{62}\Omega_c^{46} + \rho_{72}\Omega_c^{47} + \rho_{22}\Omega_p - \rho_{44}\Omega_p\big)\Big)$$

$$\dot\rho_{43} = \frac{1}{2}\left(-\gamma_3\rho_{43} - \gamma_4\rho_{43} + 2i\left(\delta_1\rho_{43} + \rho_{53}\Omega_c^{45} + \rho_{63}\Omega_c^{46} + \rho_{73}\Omega_c^{47} + \rho_{23}\Omega_p\right)\right)$$

$$\dot\rho_{44} = i\left((\rho_{54} - \rho_{45})\Omega_c^{45} + \Omega_c^{46}(\rho_{64} - \rho_{46}) + \Omega_c^{47}(\rho_{74} - \rho_{47}) + \Omega_p(\rho_{24} - \rho_{42})\right) \\ + \gamma_5\rho_{55} + \gamma_6\rho_{66} - \gamma_4\rho_{44}$$

$$\dot\rho_{45} = -\frac{1}{2}(\gamma_4 + \gamma_5)\rho_{45} - i(\delta_2 + \delta_3 + \Delta_c)\rho_{45} - i\rho_{44}\Omega_c^{45} + i\rho_{55}\Omega_c^{45} + i\rho_{65}\Omega_c^{46} \\ + i\rho_{75}\Omega_c^{47} + i\rho_{25}\Omega_p$$

$$\dot\rho_{46} = \frac{1}{2}\Big(-\gamma_4\rho_{46} - \gamma_6\rho_{46} \\ - 2i\big(\delta_3\rho_{46} + \Delta_c\rho_{46} - \rho_{56}\Omega_c^{45} + \rho_{44}\Omega_c^{46} - \rho_{66}\Omega_c^{46} - \rho_{76}\Omega_c^{47} - \rho_{26}\Omega_p\big)\Big)$$

$$\dot\rho_{47} = \frac{1}{2}\Big(-\gamma_4\rho_{47} - \gamma_7\rho_{47} \\ - 2i\big(\Delta_c\rho_{47} - \rho_{57}\Omega_c^{45} - \rho_{67}\Omega_c^{46} + \rho_{44}\Omega_c^{47} - \rho_{77}\Omega_c^{47} - \rho_{27}\Omega_p\big)\Big)$$

$$\dot\rho_{51} = -\frac{1}{2}(\Gamma_{12} + \gamma_5)\rho_{51} + i\rho_{41}\Omega_c^{45} + i\rho_{51}(\delta_2 + \delta_3 + \Delta_c + \Delta_p + \omega_{\text{HFS}})$$

$$\dot{\rho}_{52} = -\frac{1}{2}(\Gamma_{12} + \gamma_5)\rho_{52} + i(\delta_2 + \delta_3 + \Delta_c + \Delta_p)\rho_{52} + i\rho_{42}\Omega_c^{45} - i\rho_{54}\Omega_p$$

$$\dot{\rho}_{53} = -\frac{1}{2}(\gamma_3 + \gamma_5)\rho_{53} + i(\delta_1 + \delta_2 + \delta_3 + \Delta_c)\rho_{53} + i\rho_{43}\Omega_c^{45}$$

$$\dot{\rho}_{54} = -\frac{1}{2}(\gamma_4 + \gamma_5)\rho_{54} \\ + i(\delta_2\rho_{54} + \delta_3\rho_{54} + \Delta_c\rho_{54} + \rho_{44}\Omega_c^{45} - \rho_{55}\Omega_c^{45} - \rho_{56}\Omega_c^{46} - \rho_{57}\Omega_c^{47} \\ - \rho_{52}\Omega_p)$$

$$\dot{\rho}_{55} = -\gamma_5\rho_{55} + i(\rho_{45} - \rho_{54})\Omega_c^{45}$$

$$\dot{\rho}_{56} = \frac{1}{2}\left(-\gamma_5\rho_{56} - \gamma_6\rho_{56} + 2i(\delta_2\rho_{56} + \rho_{46}\Omega_c^{45} - \rho_{54}\Omega_c^{46})\right)$$

$$\dot{\rho}_{57} = -\frac{1}{2}(\gamma_5 + \gamma_7)\rho_{57} + i(\delta_2 + \delta_3)\rho_{57} + i\rho_{47}\Omega_c^{45} - i\rho_{54}\Omega_c^{47}$$

$$\dot{\rho}_{61} = -\frac{1}{2}(\Gamma_{12} + \gamma_6)\rho_{61} + i\rho_{41}\Omega_c^{46} + i\rho_{61}(\delta_3 + \Delta_c + \Delta_p + \omega_{HFS})$$

$$\dot{\rho}_{62} = -\frac{1}{2}(\Gamma_{12} + \gamma_6)\rho_{62} + i(\delta_3 + \Delta_c + \Delta_p)\rho_{62} + i\rho_{42}\Omega_c^{46} - i\rho_{64}\Omega_p$$

$$\dot{\rho}_{63} = -\frac{1}{2}(\gamma_3 + \gamma_6)\rho_{63} + i(\delta_1 + \delta_3 + \Delta_c)\rho_{63} + i\rho_{43}\Omega_c^{46}$$

$$\dot{\rho}_{64} = -\frac{1}{2}(\gamma_4 + \gamma_6)\rho_{64} \\ + i(\delta_3\rho_{64} + \Delta_c\rho_{64} - \rho_{65}\Omega_c^{45} + \rho_{44}\Omega_c^{46} - \rho_{66}\Omega_c^{46} - \rho_{67}\Omega_c^{47} - \rho_{62}\Omega_p)$$

$$\dot{\rho}_{65} = \frac{1}{2}\left(-\gamma_5\rho_{65} - \gamma_6\rho_{65} - 2i(\delta_2\rho_{65} + \rho_{64}\Omega_c^{45} - \rho_{45}\Omega_c^{46})\right)$$

$$\dot{\rho}_{66} = -\gamma_6\rho_{66} + i(\rho_{46} - \rho_{64})\Omega_c^{46}$$

$$\dot{\rho}_{67} = \frac{1}{2}\left(-\gamma_6\rho_{67} - \gamma_7\rho_{67} + 2i(\delta_3\rho_{67} + \rho_{47}\Omega_c^{46} - \rho_{64}\Omega_c^{47})\right)$$

$$\dot{\rho}_{71} = -\frac{1}{2}(\Gamma_{12} + \gamma_7)\rho_{71} + i\rho_{41}\Omega_c^{47} + i\rho_{71}(\Delta_c + \Delta_p + \omega_{HFS})$$

$$\dot{\rho}_{72} = -\frac{1}{2}(\Gamma_{12} + \gamma_7)\rho_{72} + i(\Delta_c + \Delta_p)\rho_{72} + i\rho_{42}\Omega_c^{47} - i\rho_{74}\Omega_p$$

$$\dot{\rho}_{73} = -\frac{1}{2}(\gamma_3 + \gamma_7)\rho_{73} + i(\delta_1 + \Delta_c)\rho_{73} + i\rho_{43}\Omega_c^{47}$$

$$\dot{\rho}_{74} = \frac{1}{2}\left(-\gamma_4\rho_{74} - \gamma_7\rho_{74}\right.$$
$$\left. + 2i\left(\Delta_c\rho_{74} - \rho_{75}\Omega_c^{45} - \rho_{76}\Omega_c^{46} + \rho_{44}\Omega_c^{47} - \rho_{77}\Omega_c^{47} - \rho_{72}\Omega_p\right)\right)$$

$$\dot{\rho}_{75} = \frac{1}{2}\left(-\gamma_5\rho_{75} - \gamma_7\rho_{75} - 2i(\delta_2\rho_{75} + \delta_3\rho_{75} + \rho_{74}\Omega_c^{45} - \rho_{45}\Omega_c^{47})\right)$$

$$\dot{\rho}_{76} = \frac{1}{2}\left(-\gamma_6\rho_{76} - \gamma_7\rho_{76} - 2i(\delta_3\rho_{76} + \rho_{74}\Omega_c^{46} - \rho_{46}\Omega_c^{47})\right)$$

$$\dot{\rho}_{77} = -\gamma_7\rho_{77} + i(\rho_{47} - \rho_{74})\Omega_c^{47}$$

Where $\gamma_i$ is the spontaneous decay from level $|i\rangle$, $\Omega_c^{ij}$ is the coupling Rabi frequency from level $|i\rangle \to |j\rangle$, $\Omega_p^{ij}$ is the coupling Rabi frequency from level $|i\rangle \to |j\rangle$, $\Gamma_{ij} = \frac{\gamma_i + \gamma_j}{2}$

The DROP signal is proportional to the absorption of the probe beam and is expressed as

$$\text{DROP} = \int_0^\infty Im[\rho_{24}(v)]dz \int_{-\infty}^\infty W(v)\,dv$$

where $W(v)$ is the Boltzmann velocity distribution from our thin cell. The fluorescence emitted from the $4D_{5/2}$ manifold is given by

$$\text{Fluorescence} = \sum_{i=5}^{7} \int_0^\infty \rho_{ii} dz \int_{-\infty}^\infty W(v)\,dv$$

Where we are summing over the population density matrix elements ($\rho_{ii}, i \in 5,6,7$). Finally, the FDROP signal, corresponding to the fluorescence induced by the probe beam, is calculated as

$$\text{FDROP} = \sum_{i=3}^{5} \int_0^\infty \rho_{ii} dz \int_{-\infty}^\infty W(v)dv$$

where the summation includes the relevant intermediate and excited states. The Boltzmann velocity distribution is equals $W(v) = (1/u\sqrt{\pi})\exp(-v^2/u^2)$ here u is the most probable velocity $u = \sqrt{2k_BT/m}$ where m is the mass of the atoms and T their absolute temperature.

## S2) Coefficient strength factor

Transition strength factor $S_{FF'}$ [1] for the $5P_{3/2} \rightarrow 4D_{5/2}$ transition

$$S_{FF'} = (2F'+1)(2J+1) \begin{Bmatrix} J & J' & 1 \\ F' & F & I \end{Bmatrix}^2$$

| F | F' | $S_{FF'}$ |
|---|----|-----------|
| 4 | 3  | 0.02      |
| 4 | 4  | 0.16      |
| 4 | 5  | 0.82      |

## S3) Microfabrication of Multichannel Ultrathin Rubidium Vapor Cells

The vapor cell is fabricated from two glass substrates bonded by anodic bonding. The first substrate is a 50.8 mm diameter, 1.3 mm thick Borofloat 33 glass wafer patterned to define the cell geometry. A 250 nm amorphous silicon (a-Si) layer is deposited by PECVD and patterned using direct laser-writing lithography. The a-Si is etched by RIE, followed by wet etching of the exposed glass in $HF:H_2O$ (1:10) at an etch rate of ~6 µm/h. After stripping the resist and a-Si, this process is repeated three times to form three channels with thicknesses of 8 µm, 5 µm, and 0.5 µm. The final structure consists of three 40 mm-long, 1 mm-wide channels oriented 120° apart with a common intersection. To enable anodic bonding, a second 250 nm a-Si layer is deposited and patterned using a PECVD $SiO_2$ hard mask, ensuring that the bonding regions remain outside the channels. In parallel, a second glass substrate is prepared by drilling a 4 mm through-hole, thinning the substrate to ~2 mm, and attaching a glass filling stem. The surface is polished to a roughness below 1.2 nm. The two substrates are anodically bonded at 350 °C under 700 V for 40 min using graphite interlayers. After bonding, the cell is baked under turbo-pumped vacuum for 48 h to reach pressures of $10^{-7}$–$10^{-8}$ Torr. A droplet of natural rubidium is then introduced by condensing Rb vapor produced via reduction of RbCl with calcium, and the cell is subsequently sealed. For more information see [2]

[1] D. Steck, Rubidium 85 D line data, Physics (College Park. Md). 31 (2001).